\documentclass[a4paper,11pt]{article}
\usepackage{pos}
\usepackage{wrapfig}
\usepackage{subcaption}

\newcounter{daggerfootnote}
\newcommand*{\daggerfootnote}[1]{%
    \setcounter{daggerfootnote}{\value{footnote}}%
    \renewcommand*{\thefootnote}{\fnsymbol{footnote}}%
    \footnote[2]{#1}%
    \setcounter{footnote}{\value{daggerfootnote}}%
    \renewcommand*{\thefootnote}{\arabic{footnote}}%
    }

\title{Identifying muon rings in VERITAS data using convolutional neural networks trained on images classified with Muon Hunters 2}
 \ShortTitle{Identification of Muon Rings with a MH2-trained CNN}

\author*[a]{Kevin Flanagan}
\author[a]{and John Quinn}
\author[]{for the VERITAS Collaboration\daggerfootnote{a complete list of authors can be found at the end of the proceedings}}
\author[b]{Darryl Wright}
\author[c]{Hugh Dickinson}
\author[b]{Patrick D. Wilcox}
\author[b]{Michael Laraia}
\author[c]{Stephen Serjeant}

\affiliation[a]{University College Dublin}

\affiliation[b]{University of Minnesota}

\affiliation[c]{The Open University, UK}

\emailAdd{kevin.flanagan@ucdconnect.ie}

\abstract{Muons from extensive air showers appear as rings in images taken with imaging atmospheric Cherenkov telescopes, such as VERITAS. These muon-ring images are used for the calibration of the VERITAS telescopes, however the calibration accuracy can be improved with a more efficient muon-identification algorithm. Convolutional neural networks (CNNs) are used in many state-of-the-art image-recognition systems and are ideal for muon image identification, once trained on a suitable dataset with labels for muon images. However, by training a CNN on a dataset labelled by existing algorithms, the performance of the CNN would be limited by the suboptimal muon-identification efficiency of the original algorithms. Muon Hunters 2 is a citizen science project that asks users to label grids of VERITAS telescope images, stating which images contain muon rings. Each image is labelled 10 times by independent volunteers, and the votes are aggregated and used to assign a `muon' or `non-muon' label to the corresponding image. An analysis was performed using an expert-labelled dataset in order to determine the optimal vote percentage cut-offs for assigning labels to each image for CNN training. This was optimised so as to identify as many muon images as possible while avoiding false positives. The performance of this model greatly improves on existing muon identification algorithms, identifying approximately 30 times the number of muon images identified by the current algorithm implemented in VEGAS (VERITAS Gamma-ray Analysis Suite), and roughly 2.5 times the number identified by the Hough transform method, along with significantly outperforming a CNN trained on VEGAS-labelled data.}

\FullConference{37$^{\rm{th}}$ International Cosmic Ray Conference (ICRC 2021)\\
		July 12th -- 23rd, 2021\\
		Online -- Berlin, Germany}

\begin{document}
\maketitle

\section{Introduction}
Imaging atmospheric Cherenkov telescopes (IACTs) detect high-energy gamma radiation from astrophysical sources by imaging the Cherenkov light from extensive air showers (EAS) produced when the gamma rays interact with the Earth’s atmosphere. As well as gamma rays, IACTs are also able to detect high-energy cosmic ray particles, which form a large source of background for the telescope. While most of this background is treated as noise, signal from muon particles, which are by-products of the EAS of cosmic ray particles, may be used for calibration purposes. The cone of Cherenkov light from an individual muon particle produces a ring-shaped image in an IACT. The Cherenkov opening angle (ring radius) and number of photons emitted per unit path are dependent on the energy of the muon. Therefore by determining the radius of the ring and measuring the amount of signal in that ring in the telescope, this can be compared with the amount of signal expected based on the ring size. By performing this comparison with many different muon rings, an average calibration factor for each telescope may be determined, which is used to correct for the light which is lost on the way to each telescope camera. To carry out this procedure, a large number of muon images must be identified. With advances
in machine learning and particularly in image classification through the use of convolutional neural networks (CNNs), it is possible for the rate of automatic detection of muon images to be improved dramatically over existing algorithms.

\section{VERITAS Array}
The Very Energetic Radiation Imaging Telescope Array System (VERITAS) is an array of 4 IACTs based at the Fred Lawrence Whipple Observatory in Arizona and operates in the energy range of 85 GeV to 30 TeV. The array consists of four 12 m diameter telescopes arranged in a roughly rectangular pattern, each a distance of about 100 m apart. The telescopes have Davies-Cotton segmented mirrors, each consisting of 350 individual hexagonal mirrors, with a total mirror area of 110 m$^2$. Each telescope's PMT camera has 499 pixels and a field of view of 3.5$^{\circ}$. VERITAS became operational in 2007 and is still operational currently. The array has a trigger system which requires that at least 2 telescopes simultaneously pass their own telescope-level trigger in order for each telescope in the array to record an image. This is largely to reject background events, including muon events, which typically only trigger one telescope. Many muon images contain additional Cherenkov light from the EAS from which they were produced in order to trigger the array.

\section{Methods for the Identification of Muon Images}
VEGAS is one of the two analysis packages used by VERITAS in analysing data from the telescopes \cite{VEGAS}, and it includes a muon identification algorithm. It works based on the distribution of signal in each image, determining the centroid of the signal and placing an annulus at the average radius (known as muon radius) of signal out from the centroid. The annulus has a width of 3$\sigma$
of the signal distance to the mean radius. If more than 70\% of the signal in the image is within the annulus it is classified as a muon image. While this method can identify full muon rings, there are many rings which it doesn't identify, such as partial muon rings, which occur when the muon's impact location is outside or near the edge of the telescope. This makes some photons miss the telescope and thus causes part of the ring to not show in the camera. VEGAS also doesn't identify truncated rings, where the ring is cut off by the edge of the camera due to the muon's incident angle with respect to the telescope's optical axis being greater than 1.75$^{\circ}$, nor ring images which contain signal other than the ring. These last two are currently not usable for calibration, however. 

An algorithm called the Hough transform may also be used for muon identification \cite{hough}. This algorithm is used to parameterise assumed shapes in digital images. This has been used to parameterise assumed circles in VERITAS images and then have cuts on these parameters applied to determine the presence or absence of a muon ring. This has been shown to achieve better results than the previously described VEGAS algorithm.

Convolutional neural networks (CNNs) \cite{CNN} may also be used to identify muon-ring images. CNNs are a type of machine learning algorithm commonly used in image classification tasks, where they have shown very high performance. They take an image as input and pass it through convolutional layers, pooling layers and finally fully-connected layers to produce a final output. In a binary classification problem such as muon identification, there is a single output, taking a value between 0 and 1. This number indicates the model's prediction of how likely an image is to belong to one of these classes as opposed to the other. CNNs learn to classify images by being trained on a large dataset of relevant labelled images. Each image in a training set is labelled either with a 0 or a 1, depending on its true label. The CNN predicts a value between 0 and 1 for each image, and based on how close its predictions are to the true labels, the parameters of the CNN are adjusted using an algorithm known as gradient descent. Following training, a cut-off is chosen in order to turn the outputs into classifications, typically taking  a value of 0.5. 
Provided with a large labelled dataset of muon and non-muon images, a CNN may be trained to perform muon ring identification. 

CNN classification of VERITAS muon images has been done previously, 
first using a CNN trained on VEGAS-labelled data \cite{feng1}, and then using data from the original \textit{Muon Hunter} \cite{bird} project to train a CNN \cite{feng2}. 
However, the first of these used VEGAS-labelled data to both train and test the model, which didn't test the generalisability of the model. The second chose a single vote-percentage cut-off (discussed further in section \ref{MH2_sec}) for labelling \textit{Muon Hunter} images as `muon' or `non-muon', without an analysis of what the best cut-offs would be. In this paper we both analyse model performances to determine optimised vote-percentage cut-offs for the task and test the models on a more independent dataset to better understand the generalisability of the model.

\section{Training and Validation Dataset: Muon Hunters 2} \label{MH2_sec}
Images labelled using the VEGAS algorithm may be used to create a labelled dataset for training a CNN. However, in order to ensure that the CNN is able to identify muon images which the VEGAS algorithm cannot identify, it is better to have a labelled dataset which includes these types of images also in the muon data. One way of obtaining such a dataset with more accurate labels is via human volunteers. For this reason the \textit{Muon Hunters 2} (MH2) \cite{Laraia} project was created.

\begin{wrapfigure}{r}{0.65\textwidth}
    \centering
	\includegraphics[width=0.65\textwidth]{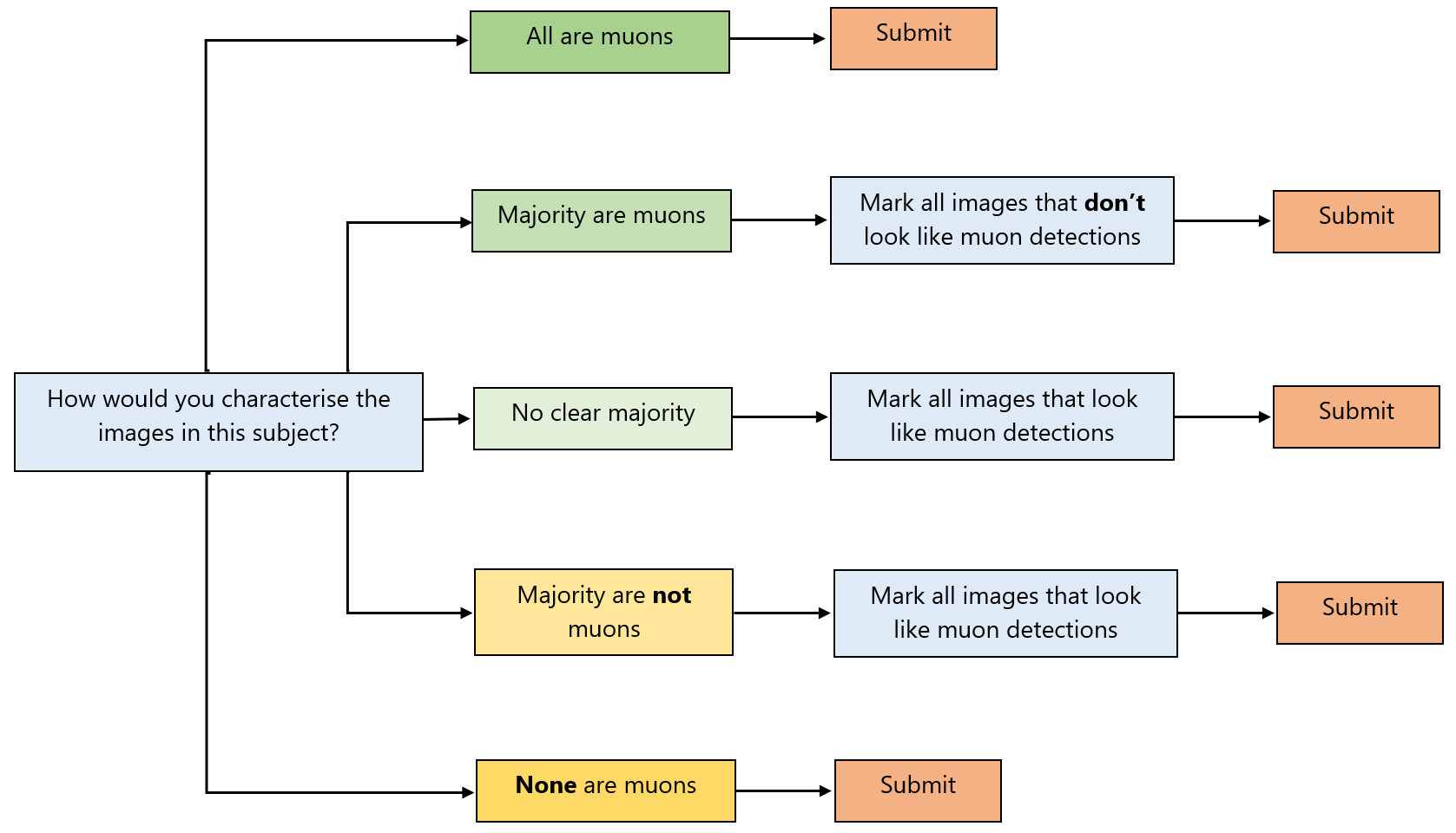}
	\caption{Schematic of the possible scenarios a volunteer will encounter in the \textit{Muon Hunters 2} project.}
	\label{MH2_options}
\end{wrapfigure}

Muon Hunters 2 is a citizen science project for gathering muon labels for VERITAS images, hosted on Zooniverse.org. As a follow-on from the original Muon Hunter project \cite{bird}, Muon Hunters 2 was launched in March of 2019 with the aim of making the identification of muon rings more efficient. In this project, a 6 $\times$ 6 grid of images is presented, and volunteers are asked to classify all of these images simultaneously by clicking on all images belonging to the minority class in that grid (muon or non-muon). This works to reduce the number of clicks required from volunteers. Figure \ref{MH2_options} displays all possible scenarios.
Volunteers are asked to identify images which contain muon rings of any variety, be they full rings, partial rings or truncated rings, and also regardless of whether there is any signal other than the ring present in the image.
\indent Each grid of images 
is retired once it has been classified by 10 independent volunteers. 
2.9 million images from real VERITAS data were included as part of the MH2 project, along with 95,000 images generated from CORSIKA simulations. For this study, a total of 605,057 labelled Muon Hunters 2 images were available and were used. The volunteer votes were used to divide the data into muon and non-muon categories for the purposes of training a CNN. Vote percentage cut-offs were used to assign images muon and non-muon labels, such that an image was assigned a muon label if voted a muon by at least X$\%$ of volunteers and non-muon if voted a muon by at most Y$\%$. These cut-off values (X$\%$ and Y$\%$) were determined through testing on a different, independently labelled dataset (process explained in section \ref{vote_cut}). The optimal cut-offs which were used to generate the labelled dataset for this study were a 20$\%$ muon and 0$\%$ non-muon cut-off. A dataset of 10,000 muon images and 10,000 non-muon images was selected and split into training, validation and test sets in a 65$\%$ : 17.5$\%$ : 17.5$\%$ split.

A VEGAS-labelled dataset was also produced, containing 10,000 muon images and 10,000 non-muon images. The same 65$\%$ : 17.5$\%$ : 17.5$\%$ split was also applied to this data. VEGAS-labelled images had an extra condition that the muon radius $r$ must be greater than 0.5$^{\circ}$ in order to increase the purity of the images, as it's found that the algorithm does not perform well on images with smaller muon radii than this.

Each image used as input to the CNN was a jpeg generated from the charge produced in the telescope camera. This was different to the oversampling approach used by Feng et al. \cite{feng1, feng2} as that method led to a stretching of the image which we wished to avoid.

\section{CNN Training} \label{training}
Both the VEGAS-labelled and MH2 datasets were used to train a CNN for the identification of muon images. The Python deep learning library \textit{keras}\footnote{\url{https://keras.io}} with a \textit{Tensorflow}\footnote{\url{https://www.tensorflow.org}} backend was used to implement the CNN model. A VGG16 \cite{vgg16} architecture was used for the CNN. This was pretrained on the ImageNet \cite{imagenet} dataset, which is a large dataset of labelled images often used to benchmark and pre-train image classification models. The same training parameters were used for each dataset. Mini-batch stochastic gradient descent with momentum was used as the optimisation algorithm. A learning rate of 0.005 and a momentum value of 0.9 were chosen, with a batch size of 64. A binary-cross-entropy cost function was used. Training was carried out for 50 epochs (an epoch being one full cycle of training on the entire dataset). The training and validation set cost values were tracked during training. As the validation cost measures the model's ability to generalise to unseen data, the model weights which minimised this value were saved as the optimised weights.

\section{Determination of Optimal Vote-Percentage Cut-Offs} \label{vote_cut}
As there are many ways to divide the MH2 data based on volunteer votes, it was necessary to determine the best way to do this for training a CNN for muon identification. A large number of different datasets were created using different pairs of vote-percentage cut-offs and models were trained on each dataset. Each dataset consisted of 17,200 images, split evenly between images with muon labels and non-muon labels. This dataset size was chosen in order to make each dataset the same size, as this was the largest possible for all datasets produced via the vote-percentage cut-offs. These datasets were further divided into training (10,320 images), validation (3,440 images) and test sets (3,440 images) each with equal numbers of muon and non-muon images. The performance of each model was tested using an independent test dataset with expert labels produced by one of the authors of this study. This test set contained 2,000 muon-labelled images and 2,000 non-muon-labelled images. It was found that overall accuracy (fraction of images classified correctly) on the independent test set was maximised by training the CNN on a dataset generated with (20,0) vote-percentage cut-offs. This means that a cut-off of 20$\%$ was applied for labelling muon images (images were labelled as muon images if their muon vote-percentage was 20$\%$ or higher) and a cut-off of 0$\%$ was applied for labelling non-muon images (images were labelled as non-muon if their muon vote-percentage was 0$\%$). This model achieved a test set accuracy of 97.1\%, with 43 false positives and 73 false negatives. It was also found that by adjusting the classification boundary on the (20,0) model output (from the default 0.5, where an output $\geq$ 0.5 gives a non-muon classification and an output $<$ 0.5 gives a muon classification) that this model could achieve 0 false positives while retaining a higher overall accuracy than other models which achieved 0 false positives. This is an important factor in choosing this model as we particularly wished to avoid false positives.

The (20,0) model achieving the best performance on the test data can likely be explained by aspects of volunteer labelling ability and the presence of ambiguous muon images.
It is generally easier for volunteers to label non-muon images correctly rather than muon images due to the presence of more difficult-to-identify muon images which volunteers may miss. These can contain partial and truncated rings, sometimes partially obscured by other signal in the image. This leads to many muon images only being labelled as containing a muon ring by volunteers a small number of times. This means that higher vote-percentage cut-offs for non-muons would lead to non-muon datasets that are contaminated by some of these more ambiguous muon images, which would confuse any model. Therefore a vote-percentage cut-off of 0\% produces what is closest to a pure set of non-muon images obtainable in this manner. 
The high performance of the low (20$\%$) muon cut-off likely comes down to increasing how representative the dataset is. The higher the muon vote-percentage used to generate the muon dataset, the less representative the dataset will be of all of the types of muon images that exist, as the muon images that are identified correctly by volunteers a large amount of the time are likely to be very clear-cut muon images.
Therefore the model will only learn to identify these more obvious images. If a smaller vote-percentage 
\begin{wrapfigure}{r}{0.5\textwidth}
  \centering
  \includegraphics[width=0.5\textwidth]{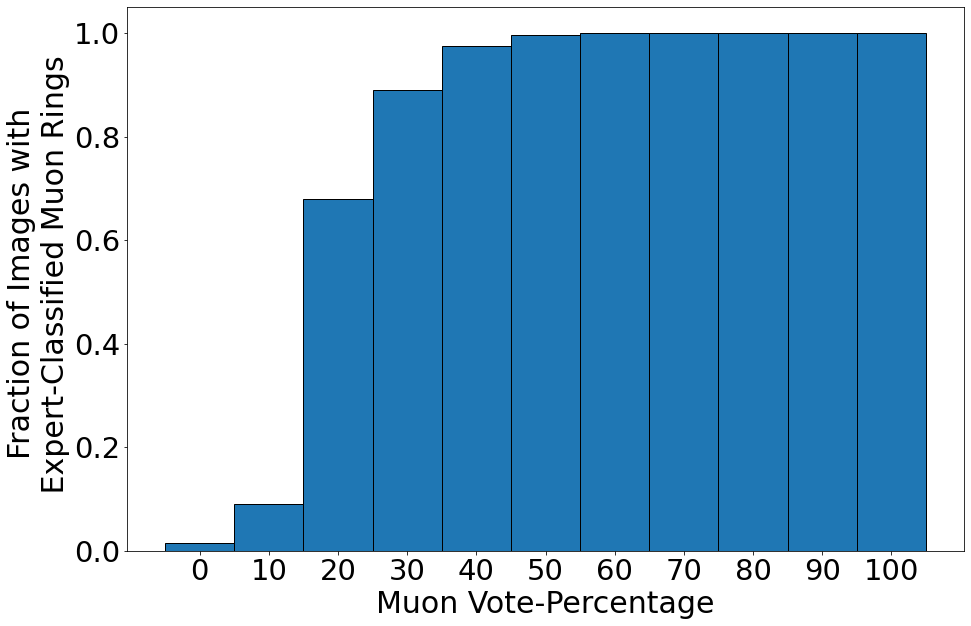}
  \caption{Fraction of images at each muon vote-percentage which contain expert-classified muon rings, based on 1,000 random images at each muon vote-percentage. Note the large increase at 20$\%$.}
  \label{per}
\end{wrapfigure}
is used to generate the muon dataset, the images become more representative of all muon image types, so the model will be better able to identify all muon images. While the level of contamination from non-muon images in the muon set also increases, it appears that the benefits of the increase in representativity of the data outweigh this. Figure \ref{per} highlights how even at the low vote-percentage of 20\% roughly 7 out of 10 images contain muon rings according to expert labels, while the datasets become fully composed of expert-labelled muons from 60\% and above. This displays the prevalence of more ambiguous muon images which volunteers found harder to identify at low vote-percentages.

\section{Final Model Performances}

The performances of the models trained on VEGAS-labelled and MH2-labelled data were evaluated on a final 10,000 image (50\% muon, 50\% non-muon) test dataset which again had expert labels produced by one of the authors. The results are displayed in Table \ref{table:results}. 

\begin{table}[h!]
\centering
\begin{tabular}{|c|c|c|}
\hline
            \textbf{Model}    & \textbf{Accuracy (\%)}                                                                 & \textbf{Maximum number of pure muons} \\ \hline
VEGAS  & 69.9   & 551 (11\%) \ \textit{(boundary = 0.024)}             \\ \hline
MH2  & 95.7   & 3892 (78\%) \ \textit{(boundary = 0.064)}            \\ \hline
\end{tabular}
\caption{Summary of VEGAS-trained and MH2-trained model performances on the expert-labelled test set.}
\label{table:results}
\end{table}
The VEGAS-trained model achieves a moderate accuracy score, while the MH2-trained model achieves a significantly higher score, correctly classifying over 95\% of images. The most important metric to consider for these models is the `maximum number of pure muons' which the model is able to retrieve. The number is obtained by adjusting the decision boundary of the output (increasing the confidence which the model must have to make a muon classification) until the model produces no false positive muon classifications on the test dataset. This represents the approximate portion of muon images which the model would be able to reliably detect without producing false positives. If one of these models was implemented in the VERITAS telescope, the decision boundary on the model would be adjusted to at least this minimum confidence level in order to obtain pure sets of muon detections. The MH2-trained model was able to detect a significantly larger pure set of muon images than the VEGAS-trained model, with $\sim$78$\%$ detected vs $\sim$11$\%$. 

Although a direct application and comparison was not made with the Hough Transform algorithm, a comparison with the results reported in a previous study by Tyler et al. \cite{hough} was made. In that paper, a similar analysis was performed on the percentage of muon images identified by the algorithm (muon efficiency). The Hough Transform study also used a set of images examined by eye and estimated that about 6.5\% of images in a typical observation run were muon images, which agrees very closely with this study's evaluation when creating labelled test data (also 6.5\%). This indicates that the test datasets used in each study likely have a similar make-up. Upon application of the Hough transform however, they achieved 29\% muon efficiency with no false positives, whereas the MH2 CNN in this study achieves 78\% muon efficiency with no false positives, indicating a substantial increase in ability to detect muon images over the Hough Transform method also.

The performance of the MH2-trained model was also compared with the original VEGAS algorithm.
A set of 481,819 images from a VERITAS observation run were taken and each muon detection method was applied to these images. The CNN models were each applied with the muon/non-muon output cut-off value which resulted in the largest pure set of muons being isolated from the expert-labelled test dataset. The number of muon images identified by these models was compared with the number detected by the VEGAS algorithm (with a $r$ $\geq$ 0.5$^{\circ}$ cut-off applied). The numbers are displayed in Table \ref{table:veg}.
\begin{table}[h!]
\centering
\begin{tabular}{|c|c|}
\hline
\textbf{Identification method}                                                                 & \textbf{Number of muon images identified} \\ \hline
VEGAS algorithm                                                                       & 728                              \\ \hline
VEGAS-trained CNN & 3071                             \\ \hline
MH2-trained CNN                                                                         & 23748                            \\ \hline
\end{tabular}
\caption{Number of muon images identified by each method out of the 481,819 images in the dataset.}
\label{table:veg}
\end{table}
The VEGAS-trained CNN does show an improvement over the original algorithm, showing generalisation from the CNN training, however the MH2-trained CNN far exceeds this, identifying $\sim$32 times more muon images than the VEGAS algorithm. 

\section{Conclusion}

This study has investigated the use of CNNs for muon identification in VERITAS images. 
Tests were carried out to determine the best way of using the volunteer votes from the Muon Hunters 2 project to generate labels, and a CNN trained on these labelled images outperformed a CNN trained on VEGAS-labelled data. Both CNNs were compared with the current VEGAS muon detection algorithm, and both achieved a higher rate of detection of muon images. This was especially true for the MH2 CNN, identifying approximately 32 times the number of muon images that the VEGAS algorithm did. This was achieved with the output value cut-off of the CNN adjusted so as to eliminate false positives from the muon images identified. Certain muon images may be used for calibration of the VERITAS telescopes. However, as the MH2-trained CNN is trained to identify all types of muon images, including those which are not usable for calibration such as those with truncated rings and with signal other than the ring, some extra work is needed to extract only images which are suitable for calibration. This may be similar to what is currently done with the VEGAS algorithm where certain extra quality cuts are applied. As a significantly larger number of muon images are identified by the MH2-trained CNN overall, it should also be the case that a larger number of images suitable for calibration are identified as well.

\section*{Acknowledgements}
This research is supported by grants from the U.S. Department of Energy Office of Science, the U.S. National Science Foundation and the Smithsonian Institution, by NSERC in Canada, and by the Helmholtz Association in Germany. This research used resources provided by the Open Science Grid, which is supported by the National Science Foundation and the U.S. Department of Energy's Office of Science, and resources of the National Energy Research Scientific Computing Center (NERSC), a U.S. Department of Energy Office of Science User Facility operated under Contract No. DE-AC02-05CH11231. We acknowledge the excellent work of the technical support staff at the Fred Lawrence Whipple Observatory and at the collaborating institutions in the construction and operation of the instrument.

This publication uses data generated via the Zooniverse.org platform, development of which is funded by generous support, including a Global Impact Award from Google, and by a grant from the Alfred P. Sloan Foundation.

This project has been partly supported by ESCAPE - The European Science Cluster of Astronomy $\&$ Particle Physics ESFRI Research Infrastructures, which has received funding from the European Union’s Horizon 2020 research and innovation programme under Grant Agreement no. 824064.

\clearpage \section*{Full Authors List: \Coll\ Collaboration}

\scriptsize
\noindent
C.~B.~Adams$^{1}$,
A.~Archer$^{2}$,
W.~Benbow$^{3}$,
A.~Brill$^{1}$,
J.~H.~Buckley$^{4}$,
M.~Capasso$^{5}$,
J.~L.~Christiansen$^{6}$,
A.~J.~Chromey$^{7}$, 
M.~Errando$^{4}$,
A.~Falcone$^{8}$,
K.~A.~Farrell$^{9}$,
Q.~Feng$^{5}$,
G.~M.~Foote$^{10}$,
L.~Fortson$^{11}$,
A.~Furniss$^{12}$,
A.~Gent$^{13}$,
G.~H.~Gillanders$^{14}$,
C.~Giuri$^{15}$,
O.~Gueta$^{15}$,
D.~Hanna$^{16}$,
O.~Hervet$^{17}$,
J.~Holder$^{10}$,
B.~Hona$^{18}$,
T.~B.~Humensky$^{1}$,
W.~Jin$^{19}$,
P.~Kaaret$^{20}$,
M.~Kertzman$^{2}$,
T.~K.~Kleiner$^{15}$,
S.~Kumar$^{16}$,
M.~J.~Lang$^{14}$,
M.~Lundy$^{16}$,
G.~Maier$^{15}$,
C.~E~McGrath$^{9}$,
P.~Moriarty$^{14}$,
R.~Mukherjee$^{5}$,
D.~Nieto$^{21}$,
M.~Nievas-Rosillo$^{15}$,
S.~O'Brien$^{16}$,
R.~A.~Ong$^{22}$,
A.~N.~Otte$^{13}$,
S.~R. Patel$^{15}$,
S.~Patel$^{20}$,
K.~Pfrang$^{15}$,
M.~Pohl$^{23,15}$,
R.~R.~Prado$^{15}$,
E.~Pueschel$^{15}$,
J.~Quinn$^{9}$,
K.~Ragan$^{16}$,
P.~T.~Reynolds$^{24}$,
D.~Ribeiro$^{1}$,
E.~Roache$^{3}$,
J.~L.~Ryan$^{22}$,
I.~Sadeh$^{15}$,
M.~Santander$^{19}$,
G.~H.~Sembroski$^{25}$,
R.~Shang$^{22}$,
D.~Tak$^{15}$,
V.~V.~Vassiliev$^{22}$,
A.~Weinstein$^{7}$,
D.~A.~Williams$^{17}$,
and 
T.~J.~Williamson$^{10}$\\
\noindent
$^1${Physics Department, Columbia University, New York, NY 10027, USA}
$^{2}${Department of Physics and Astronomy, DePauw University, Greencastle, IN 46135-0037, USA}
$^3${Center for Astrophysics $|$ Harvard \& Smithsonian, Cambridge, MA 02138, USA}
$^4${Department of Physics, Washington University, St. Louis, MO 63130, USA}
$^5${Department of Physics and Astronomy, Barnard College, Columbia University, NY 10027, USA}
$^6${Physics Department, California Polytechnic State University, San Luis Obispo, CA 94307, USA} 
$^7${Department of Physics and Astronomy, Iowa State University, Ames, IA 50011, USA}
$^8${Department of Astronomy and Astrophysics, 525 Davey Lab, Pennsylvania State University, University Park, PA 16802, USA}
$^9${School of Physics, University College Dublin, Belfield, Dublin 4, Ireland}
$^{10}${Department of Physics and Astronomy and the Bartol Research Institute, University of Delaware, Newark, DE 19716, USA}
$^{11}${School of Physics and Astronomy, University of Minnesota, Minneapolis, MN 55455, USA}
$^{12}${Department of Physics, California State University - East Bay, Hayward, CA 94542, USA}
$^{13}${School of Physics and Center for Relativistic Astrophysics, Georgia Institute of Technology, 837 State Street NW, Atlanta, GA 30332-0430}
$^{14}${School of Physics, National University of Ireland Galway, University Road, Galway, Ireland}
$^{15}${DESY, Platanenallee 6, 15738 Zeuthen, Germany}
$^{16}${Physics Department, McGill University, Montreal, QC H3A 2T8, Canada}
$^{17}${Santa Cruz Institute for Particle Physics and Department of Physics, University of California, Santa Cruz, CA 95064, USA}
$^{18}${Department of Physics and Astronomy, University of Utah, Salt Lake City, UT 84112, USA}
$^{19}${Department of Physics and Astronomy, University of Alabama, Tuscaloosa, AL 35487, USA}
$^{20}${Department of Physics and Astronomy, University of Iowa, Van Allen Hall, Iowa City, IA 52242, USA}
$^{21}${Institute of Particle and Cosmos Physics, Universidad Complutense de Madrid, 28040 Madrid, Spain}
$^{22}${Department of Physics and Astronomy, University of California, Los Angeles, CA 90095, USA}
$^{23}${Institute of Physics and Astronomy, University of Potsdam, 14476 Potsdam-Golm, Germany}
$^{24}${Department of Physical Sciences, Munster Technological University, Bishopstown, Cork, T12 P928, Ireland}
$^{25}${Department of Physics and Astronomy, Purdue University, West Lafayette, IN 47907, USA}

%
%
%

\end{document}